\documentstyle[prd,aps,preprint,epsfig]{revtex}

\begin{document}

\tightenlines

\preprint{\parbox[b]{1in}{
\hbox{\tt PNUTP-01/A01}}}

\draft

\title{Aspects of Color Superconductivity\footnote{
A Lecture given at the 40th Cracow Jubilee School of Theoretical
Physics, June 3 - 11, Zakopane, Poland, and partially at Cosmo 2000,
Sep. 4 - 8, Cheju, Korea.}
}

\author{
Deog Ki Hong\footnote{E-mail: dkhong@pnu.edu},
 }

\address{Department of Physics, Pusan National University,
Pusan, Korea}

\maketitle

\begin{abstract}
I discuss some aspects of recent developments in color
superconductivity in high density quark matter.
I calculate the Cooper pair gap and the critical points
at high density, where magnetic gluons are not screened.
The ground state of high density QCD with three light flavors
is shown to be a color-flavor locking state, which can be mapped
into the low-density hadronic phase. The meson mass at the CFL
superconductor is also calculated. The CFL color superconductor
is bosonized, where the Fermi sea is identified as a $Q$-matter and
the gapped quarks as topological excitations, called
superqualitons, of mesons. Finally, as an application of
color supercoductivity, I discuss the neutrino interactions
in the CFL color superconductor.
\end{abstract}

\section{Introduction}
Matter exhibits several different phases, as shown in Fig.~1,
depending on external parameters. At temperature,
higher than the deconfinement temperature
($T>100$~MeV), quarks confined in the nucleons get
liberated and matter becomes a quark-gluon plasma,
as happened in the very early universe.
Similarly, also at extremely high density, where the Fermi momentum
of nucleons in matter is larger than $1$ GeV or so as in the core
of compact stars like neutron stars, the wavefunction of quarks
in nucleons will overlap with that of quarks in other nucleons
due to asymptotic freedom.
At such high density, quarks are no longer confined in nucleons
and thus the nuclear matter will become a quark matter,
where rather weakly interacting quarks move
around~\cite{Collins:1975ky}.
\begin{center}
\begin{figure}[htb]
 \begin{minipage}{100mm}
 \begin{tabular}{c}
\epsfig{file=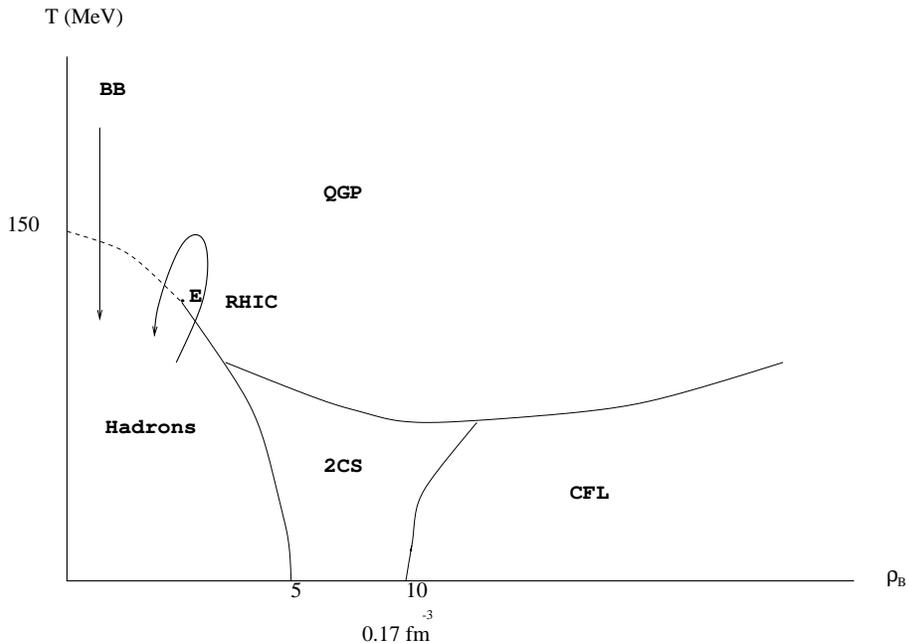,width=12cm}
 \end{tabular}
  \end{minipage}
\caption{A schematic phase diagram of matter.}
\label{diag}
\end{figure}
\end{center}
The Fermi surface of quark matter at high density
is unstable at low temperature, a phenomenon called Cooper instability,
against forming pairs of quarks or holes,
if attraction exists
between a pair of quarks or holes with opposite momenta.
No matter how small the attraction is,
it will dominate any repulsive forces at low energy, since
the attraction between a pair of quarks or holes with opposite momenta
is a relevant operator while all repulsive forces become
irrelevant as one scales down toward the Fermi
sea~\cite{Shankar:1994pf,Polchinski:1992ed}.
It turns out that
color anti-triplet diquark condensates are energetically most
preferred among possible pairings,
including particle-hole parings or
density waves~\cite{Shuster:2000tn}.

Intense study on color superconductivity~\cite{review} shows that
superconducting quark matter has two
different phases, depending on density.~\footnote{
New phases like the LOFF phase~\cite{Alford:2000ze} or
a chiral crystal phase~\cite{Rapp:2000zd}
might exist at the intermediate density if one
includes the Fermi surface mismatch due to the difference in quark
mass.} At intermediate density,
the Cooper pair is color
anti-triplet but flavor singlet, breaking only the color symmetry
down to a subgroup, $SU(3)_c\mapsto SU(2)_c$:
\begin{eqnarray}
\left< \psi_{Li}^{a}(\vec p)\psi_{Lj}^{b}(-\vec p) \right>
&=&-\left<\psi_{Ri}^{a}(\vec p)\psi_{Rj}^{b}(-\vec p)\right>
\nonumber\\
&=&\epsilon_{ij}\epsilon^{ab3}\Delta,
\label{cs2}
\end{eqnarray}
where $i,j=1,2$ and $a,b=1,2,3$ are flavor and color
indices, respectively.
For high density where the chemical potential is larger than the
strange quark mass, $\mu>m_s$, the strange quark participates
in Cooper-pairing. At such a high density,
the Cooper-pair condensate is predicted
to take a so-called color-flavor locking (CFL) form~\cite{Alford:1999mk},
breaking not only color symmetry but also flavor symmetry maximally:
\begin{eqnarray}
\left< \psi_{Li}^{a}(\vec p)\psi_{Lj}^{b}(-\vec p) \right>
&=&-\left<\psi_{Ri}^{a}(\vec p)\psi_{Rj}^{b}(-\vec p)\right>
\nonumber\\
&=&k_1\delta_i^a\delta_j^b+k_2\delta_j^a\delta_i^b.
\label{cfl}
\end{eqnarray}
At much higher density
($\mu\gg\Lambda_{\rm QCD}$), $k_1(\equiv \Delta_0)\simeq -k_2$
and the color-flavor locking phase is shown to be
energetically preferred~\cite{Evans:2000at,Hong:2000ru,Schafer:2000fe}.

\section{Cooper pair gap and the critical points}

There are two kinds of the attractive forces that lead
to Cooper instability, depending on the density.
At the intermediate density,
where $\mu<m_s$ or $\rho\sim (5 - 10)\times 0.17~{\rm fm}^{-3}$,
the QCD interaction is approximated with four-quark interactions,
\begin{equation}
{\cal L}^{\rm eff}_{\rm QCD}\ni {g^2\over \Lambda^2}
\bar\psi\psi\bar\psi\psi+\cdots,
\end{equation}
since both electric and magnetic gluons are screened due to the
medium effect.
This short-range attraction is precisely the BCS type interaction,
which is generated in metal by the exchange of massive phonons.
The Cooper pair gap is then given by~\cite{schrieffer}
\begin{equation}
\Delta\simeq \epsilon_F\exp\left[-{\Lambda^2\over g^2p_F^2}\right],
\end{equation}
which is estimated to be $10\sim 100$ MeV, for $\Lambda$ and
$\epsilon_F$ are of the order of $\Lambda_{\rm QCD}$ and $g$ is
of the order of one at the intermediate density.
On the other hand, though electric gluons are screened in quark matter,
the magnetic gluons are not screened at high density
even at a nonperturbative level,
as argued convincingly by Son~\cite{son1}. Thus
the Cooper-pairing force at high density is long-ranged and
the gap equation is so-called the Eliashberg equation.
The (long-range) magnetic gluon exchange interaction leads to an
extra (infrared) logarithmic divergence in the gap
equation, which is in hard-dense loop (HDL) approximation
given as,
\begin{eqnarray}
\Delta(p_0)&=& {g_s^2\over 36\pi^2}\int_{-\mu}^{\mu}dq_0
{\Delta(q_0)\over \sqrt{q_0^2+\Delta^2}}
\ln\left({{\bar\Lambda}\over |p_0-q_0|}
\right)
\label{gapf}
\end{eqnarray}
where $\bar\Lambda=4\mu/\pi\cdot (\mu/M)^5e^{3/2\xi}$ with a
gauge parameter $\xi$. By solving the gap equation (\ref{gapf}),
one finds
the Cooper pair gap to be~\cite{son1,Hong:2000tn,my4,sw,pr,hs}\footnote{
Were we to take the UV cutoff of the effective theory
to be $2\mu$ instead of $\mu$, taken
in~\cite{Hong:2000tn}, we would get the usual value, $2^8$, instead
of $2^2$ for $N_f=2$ in the prefactor.}
\begin{equation}
\Delta_0=
2^{9/2}\pi^4N_f^{-5/2}e^{3\xi/2+1}\cdot{\mu\over g_s^5}
\exp\left(-{3\pi^2\over\sqrt{2}g_s}\right).
\end{equation}


Though the ground state of quark matter is a color superconductor,
one needs to know its criticality to observe color superconductivity
in the laboratory or in stellar objects.
The quark matter which might exist in the core of compact stars
like neutron stars will be in the superconducting phase
if the interior temperature of compact stars is lower than
the critical temperature and the density is higher than the
critical density. For the neutron stars, the inner temperature
is estimated to be $<0.7\,{\rm MeV}$ and the core
density is around $1.7$~${\rm fm}^{-3}$, which is ten times
higher than the normal nuclear matter density~\cite{pines}.
Since the critical temperature of BCS superconductivity is given
as~\cite{schrieffer}
\begin{equation}
T_c={1\over\pi}e^{\gamma}\Delta\simeq 0.57\Delta,
\end{equation}
the critical temperature of color superconductivity at the
intermediate density is quite large; $T_c\sim 5$-50~MeV.
In dense QCD with unscreened magnetic gluons, the critical
temperature turns out to take the BCS type
value~\cite{Brown:2000yd,Pisarski:1999av,Hong:2000ru},
$T_c\simeq 0.57\Delta$, though the pairing force is very different from
that of the BCS superconductivity.
Since the unscreened magnetic gluons give
a much bigger gap than the usual BCS type gap,
the critical temperature of color superconductvity
is quite large compared to the interior temperature of neutron stars,
regardless of the form of attractive forces.

It is instructive to derive the critical temperature
for the color superconductivity at high density where the pairing
is mediated by the unscreened magnetic gluons. We start with the
zero temperature Cooper-pair gap equation, Eq.~(\ref{gapf}).

Following the imaginary-time formalism developed by
Matsubara~\cite{matsubara},
the gap equation becomes at finite temperature  $T$
\begin{eqnarray}
\Delta(\omega_{n^{\prime}})={g_s^2\over 9\pi}T
\sum_{n=-\infty}^{+\infty}
\int{{\rm d}q\over 2\pi}{\Delta(\omega_n)\over \omega_n^2+
\Delta^2(\omega_n)+q^2}\ln\left({\bar\Lambda\over \left|
\omega_{n^{\prime}}-\omega_n\right|}\right),
\end{eqnarray}
where $\omega_n=\pi T(2n+1)$ and $q\equiv \vec v_F\cdot\vec q$.
We now use the constant
(but temperature-dependent) gap approximation,
$\Delta(\omega_n)\simeq \Delta(T)$ for all $n$.
Taking $n^{\prime}=0$ and converting the logarithm into integration,
we get
\begin{eqnarray}
\Delta(T)={g_s^2\over 18\pi}T\sum_{n=-\infty}^{+\infty}
\int{{\rm d}q\over 2\pi}\int_0^{{\bar\Lambda}^2}{\rm d}x
{\Delta(T)\over \omega_n^2+\Delta^2(T)+q^2}\cdot
{1\over x+(\omega_n-\omega_0)^2}.
\end{eqnarray}
Using the contour integral~\cite{gusynin97},
one can in fact sum up over all $n$ to get
\begin{eqnarray}
1&=&{g_s^2\over36\pi^2}\int {\rm d}q\int_0^{{\bar\Lambda}^2}
{{\rm d}x\over 2\pi i}\oint_C{{\rm d}\omega\over 1+e^{-\omega/T}}
\cdot{1\over\left[\omega^2-q^2-\Delta^2(T)\right]\left[
(\omega-i\omega_0)^2-x\right]}.
\label{tgap}
\end{eqnarray}
Since the gap vanishes at the critical temperature,
$\Delta(T_C)=0$,
we get, after performing the contour integration in Eq.~(\ref{tgap}),
\begin{eqnarray}
1&=&{g_s^2\over36\pi^2}\int{\rm d}q\int_0^{{\bar\Lambda}^2}
{\rm d}x\left\{
{(\pi T_C)^2+x-q^2\over \left[
(\pi T_C)^2+x-q^2\right]^2+(2\pi T_Cq)^2}\cdot
{\tanh\left[q/(2T_C)\right]\over 2q}\right.\nonumber\\
& &\left.\quad\quad
+{(\pi T_C)^2+q^2-x\over \left[(\pi T_C)^2+q^2-x\right]^2
+(2\pi T_C)^2x}\cdot{ \coth\left[\sqrt{x}/(2T_C)\right]
\over 2\sqrt{x}}
\right\}.
\label{ttgap}
\end{eqnarray}
At high density $\bar\Lambda\gg T_C$,
the second term in the integral in Eq.~(\ref{ttgap}) is
negligible, compared to the first term, and
integrating over $x$, we get
\begin{eqnarray}
1&=&{g_s^2\over 36\pi^2}\int_0^{\lambda_c}{\rm d}y
{\tanh y\over y}\left[
\ln\left({\lambda_c^2\over (\pi/2)^2+y^2}\right)+
O\left({y^2\over \lambda_c^2}\right)\right]\nonumber\\
&=&{g_s^2\over 36\pi^2}\left[
\int_0^1{\rm d}y{\tanh y\over y}\ln\lambda_c^2
+\int_1^{\lambda_c}{\rm d}y{\tanh y\over y}\ln{\lambda_c^2\over y^2}
+\cdots\right]\\
&\simeq&{g_s^2\over 36\pi^2}\left[\ln\left(e^A\lambda_c\right)\right]^2
\nonumber
\end{eqnarray}
where we have introduced $y\equiv q/(2T_C)$ and
$\lambda_c\equiv\bar\Lambda/(2T_C)$ and $A$ is given as
\begin{equation}
A=\int_0^1{\rm d}y{\tanh y\over y}+\int_1^{\infty}{\rm d}y
{\tanh y-1\over y}=\ln \left({4\over\pi}\right)+\gamma,
\end{equation}
where the Euler-Mascheroni constant
$\gamma\simeq0.577$. Therefore, we find the critical temperture
\begin{equation}
T_C={e^A\over2}{\bar\Lambda}\exp\left(-{6\pi\over g_s}\right).
\end{equation}
Now, one can also solve the gap equation Eq.~(\ref{gapf}) in the
same approximation used to find the critical temperature.
Taking the gap independent of the energy, we get
\begin{eqnarray}
1&\simeq&
{g_s^2\over18\pi^2}\int_0^{\bar\Lambda}{{\rm d}q_0\over \sqrt{q_0^2
+\Delta^2}}\ln\left({\bar\Lambda\over q_0}\right)\nonumber\\
&=&{g_s^2\over18\pi^2}\int_0^{\bar\lambda}{{\rm d}x\over \sqrt{x^2+1}}
\left(\ln{\bar\lambda}-\ln x\right)\\
&\simeq&{g_s^2\over36\pi^2}\left[\ln\left(2\bar\lambda\right)\right]^2,
\nonumber
\end{eqnarray}
where we have introduced $x={q_0/\Delta}$,
${\bar\lambda}\equiv {\bar\Lambda/\Delta}$, and used the fact
that the gap vanishes rapidly at energy higher than $\bar\Lambda$.
In this constant gap approximation, the gap is given as
\begin{equation}
\Delta=2\bar\Lambda\exp\left(-{6\pi\over g_s}\right),
\end{equation}
which is about $1.75~T_C$.
As comparison, we note in the BCS case, which has a contact
four-Fermi interaction with strength ${\bar g}$,
the critical temperature is given as
\begin{eqnarray}
1&=&{\bar g}\int_0^{\tilde\omega_c}{\rm d}z{\tanh z\over z}\nonumber\\
&\simeq&{\bar g}\left[\int_1^{\tilde\omega_c}{{\rm d}z\over z}
+\int_0^1{\rm d}z{\tanh z\over z}-\int_1^{\infty}{\rm d}z
{1-\tanh z\over z}\right]\\
&=&{\bar g}\ln\left(e^A\tilde\omega_c\right)\nonumber
\end{eqnarray}
where $\tilde\omega_c(\gg1)$ is determined by the Debye energy,
$\tilde\omega_c=\omega_D/(2T_C)$.
Since the gap  $\Delta=2\omega_D\exp(-1/{\bar g})$ in the BCS
superconductivity,
the ratio between the critical temperature and
the Cooper-pair gap in both the color superconductivity at high density
and the BCS superconductivity is given as $e^{\gamma}/\pi\simeq 0.57$,

At high density, antiquarks are difficult to create due to the energy
gap provided by the Fermi sea and thus it is energetically
disfavored for antiquarks to participate in condensation. But, as
the density becomes lower, one has to take into account the effect
of antiquarks. In the high density effective theory, this effect is
incorporated in the higher order operators~\cite{Hong:2000ru}.
First, we add the leading $1/\mu$ corrections to the gap equation
Eq.~(\ref{gapf}) to see how the formation of
Cooper pair changes when the density decreases.
The leading
$1/\mu$ corrections to the quark-gluon interactions are
\begin{equation}
{\cal L}_1=-{1\over2\mu}\sum_{\vec v_F}\psi^{\dagger}(\vec
v_F,x)\left(\gamma_{\perp}\cdot D\right)^2\psi(\vec v_F,x)
=-\sum_{\vec v_F}\left[\psi^{\dagger}{D_{\perp}^2\over
2\mu}\psi+g_s\psi^{\dagger}{\sigma_{\mu\nu}F^{\mu\nu}\over
4\mu}\psi\right].
\end{equation}
In the leading order in the HDL approximation,
the loop correction to the vertex is
neglected and the quark-gluon vertex is shifted by the $1/\mu$
correction as
\begin{equation}
-ig_sv_F^i\mapsto -ig_sv_F^i-ig_s{l_{\perp}^i\over\mu},
\end{equation}
where $l_i$ is the momentum carried away from quarks by gluons.
We note that since the $1/\mu$ correction to the quark-gluon
vertex does not depend on the Fermi velocity of the quark, it
generates a repulsion for quark pairs, bound by magnetic forces.
For a constant gap approximation, $\Delta(p_{\parallel})
\simeq \Delta$, the gap equation becomes in the leading order,
as $p_{\parallel}\to0$,
\begin{eqnarray}
1={g_s^2\over9\pi}\int{{\rm d}^2l_{\parallel}\over
(2\pi)^2}\left[
\ln\left({\bar\Lambda\over |l_0|}\right)-{3\over2}\right]
{1\over l_{\parallel}^2+\Delta^2}
={g_s^2\over 36\pi^2}\ln\left({\bar\Lambda\over \Delta}\right)
\left[\ln\left({\bar\Lambda\over\Delta}\right)-3\right].
\end{eqnarray}
When ${\bar\Lambda}\le e^3\Delta$, the gap due
to the long-range color magnetic interaction disappears. Since the
phase transition for color superconducting phase is believed
to be of first order~\cite{phase,shuryak1}, we may assume that the gap
has the same dependence on the chemical potential $\mu$
as the leading order. Then, the critical density for the color
superconducting phase transition is given by
\begin{equation}
1=e^3\exp\left[-{3\pi^2\over \sqrt{2}g_s(\mu_c)}\right].
\end{equation}
Therefore, if the strong interaction coupling is too strong
at the scale of the chemical potential, the gap does not form.
In other words, the chemical potential has to be bigger than a
critical value, $0.13{\rm GeV}<\mu_c<0.31{\rm GeV}$, which
is about the same  as the one estimated in the
literature~\cite{phase,shuryak1,shuryak2}.

\section{The Color-Flavor-Locking phase}
When $N_f=3$, the spin-zero component of
the condensate  becomes (flavor)
anti-triplet,
\begin{eqnarray}
\left<{\psi_L}^a_{i\alpha}(\vec v_F,x){\psi_L}^b_{j\beta}(-\vec v_F,x)
\right>&=&-\left<{\psi_R}^a_{i\alpha}(\vec v_F,x)
{\psi_R}^b_{j\beta}(-\vec v_F,x)\right>\\
&=&\epsilon_{ij}\epsilon^{abc}
\epsilon_{\alpha\beta\gamma}K_c^{\gamma}(p_F),
\end{eqnarray}
where $\psi(\vec v_F,x)$ is the quark near the Fermi surface with Fermi
velocity $\vec v_F$~\cite{Hong:2000tn,Hong:2000ru}.
Using the global color and flavor symmetry, one can
always diagonalize the spin-zero condensate as
$K_c^{\gamma}=\delta_c^{\gamma}K_{\gamma}$.
To determine the parameters, $K_u$, $K_d$, and $K_s$,
we need to minimize the vacuum energy for the
condensate. By the Cornwall-Jackiw-Tomboulis
formalism~\cite{Cornwall:1974vz},
the vacuum energy in the HDL approximation is given as
\begin{eqnarray}
V(\Delta)&=& -{\rm Tr}~\ln S^{-1}+{\rm Tr}~\ln\not\!\partial
+{\rm Tr}~(S^{-1}-\not\!\partial)S+({\rm 2PI~diagrams})
\nonumber\\
&=&{\mu^2\over4\pi}\sum_{i=1}^9\int {d^2l_{\parallel}\over
(2\pi)^2}\left[\ln\left({l_{\parallel}^2\over
l_{\parallel}^2+\Delta_{i}^2(l_{\parallel})} \right)+
{1\over2}\cdot{\Delta_{i}^2(l_{\parallel})\over
l_{\parallel}^2+\Delta_{i}^2(l_{\parallel})}\right]+h.o.,
\label{ve}
\end{eqnarray}
where $h.o.$ are the higher order terms in the HDL approximation,
containing more powers of coupling $g_s$, and
$\Delta_i$'s are the eigenvalues of color anti-symmetric
and flavor anti-symmetric $9\times9$ gap, $\Delta_{\alpha\beta}^{ab}$.
The 2PI diagrams are two-particle-irreducible vacuum diagrams. There
is only one such diagram (see Fig.~2) in the leading order
HDL approximation.
\begin{figure}[h]
\vskip 0.3in
\epsfxsize=2.0in
\centerline{\epsffile{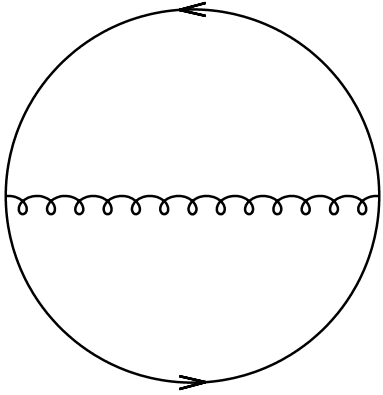}}
\caption{The 2PI vacuum energy diagram. }
 \label{2pi}
\end{figure}

Since the gap depends only on energy in the leading order,
one can easily perform the momentum integration in
(\ref{ve}) to get\footnote{
If the condensate forms, the vacuum energy due to the gluons also
depends on the gap due to the Meisner effect. But, it turns
out to be subleading, compared to the quark vacuum energy;
$V_g(\Delta)\sim M^2\Delta^2
\ln(\Delta/\mu)\sim g_s\mu^2\Delta^2$~\cite{Evans:2000at}.},
\begin{eqnarray}
V(\Delta)&=&{\mu^2\over 4\pi^2}\int_0^{\infty}
dl_0\left(-{\Delta_i^2\over
\sqrt{l_0^2}+\sqrt{l_0^2+\Delta_i^2}}+{1\over4}\cdot
{\Delta_i^2\over\sqrt{l_0^2+\Delta_i^2}}
\right)\nonumber\\
&\simeq&
-0.43 {\mu^2\over 4\pi^2}
    \sum_i\left|\Delta_i(0)\right|^2,
\end{eqnarray}
where in the second line we used an approximation that
\begin{equation}
\Delta_i(l_0)\simeq
\left\{
\begin{array}{ll}
\Delta_i(0) & \mbox{if $\left|l_0\right|<\left|\Delta_i(0)\right|$},\\
0 & \mbox{otherwise}.
\end{array}
\right.
\end{equation}
Were $\Delta_i$ independent of each other, the global minimum
should occur at $\Delta_i(0)={\rm const.}$ for all $i=1,\cdots,9$.
But, due to the global
color and flavor symmetry, only three of them are independent.
Similarly to the condensate,
the gap can be also diagonalized by the color and flavor symmetry as
\begin{equation}
\Delta^{\alpha\beta}_{ab}=\epsilon_{\alpha\beta\gamma}
\epsilon^{abc}\Delta_{\gamma}\delta^{\gamma}_c.
\end{equation}
Without loss of generality, we can take $\left|\Delta_u\right|
\ge \left|\Delta_d\right|\ge\left|\Delta_s\right|$.
Let $\Delta_d/\Delta_u=x$ and $\Delta_s/\Delta_u=y$.
Then, the vacuum energy becomes
\begin{equation}
V(\Delta)\simeq -0.43 {\mu^2\over 4\pi^2}\left|\Delta_u\right|^2f(x,y),
\end{equation}
where $f(x,y)$ is a complicate function of $-1\le x,y\le 1$
that has a maximum at $x=1=y$, $f(x,y)\le 13.4$.
Therefore, the vacuum energy has a global minimum when
$\Delta_u=\Delta_d=\Delta_s$, or in terms of the eigenvalues of the gap
\begin{equation}
\Delta_i=\Delta_u \cdot(1,1,1,-1,1,-1,1,-1,-2).
\end{equation}
Among nine quarks, $\psi_a^{\alpha}$, eight of them have (Majorana) mass
$\Delta_u$, forming an octet under $SU(3)$, and
one of them, a singlet under $SU(3)$, has $2\Delta_u$.

Since the condensate is
related to the off-diagonal component of the quark propagator at high
momentum as, suppressing the color and flavor indices,
\begin{eqnarray}
\left<\psi(\vec v_F,x)\psi(-\vec v_F,x)\right>
&\sim&\lim_{y\to x}\int {{\rm d}^4l\over (2\pi)^4}e^{il\cdot (x-y)}
{\Delta(l_{\parallel})\over l_{\parallel}^2-\Delta^2(l_{\parallel})}
\nonumber\\
&=&\lim_{y\to x}\left[\delta^2(\vec x_{\perp}-\vec y_{\perp})
{\Delta(0)\over4\pi^2|x_{\parallel}-y_{\parallel}|^{\gamma_m}}
+\cdots\right],
\end{eqnarray}
where $\gamma_m$ is the anomalous dimension of the condensate and
the ellipsis are less singular terms.
Being proportional to the gap, the condensate is
diagonalized in the basis where the gap is diagonalized.
Thus, we have shown that in the HDL approximation
the true ground state of QCD with three
massless flavors at high density is the color-flavor locking
phase,  $K_{\gamma}=K$ for all $\gamma=u,d,s$. The condensate takes
\begin{equation}
\left<{\psi_L}^a_{i\alpha}(\vec v_F,x){\psi_L}^b_{j\beta}(-\vec v_F,x)
\right>=-\left<{\psi_R}^a_{i\alpha}(\vec v_F,x)
{\psi_R}^b_{j\beta}(-\vec v_F,x)\right>
=\epsilon_{ij}\epsilon^{abI}\epsilon_{\alpha\beta I}K(p_F),
\end{equation}
breaking the color symmetry, $U(1)_{\rm em}$, the chiral symmetry,
and the baryon number symmetry.
The symmetry breaking pattern of the CFL phase
is therefore
\begin{equation}
SU(3)_c\times SU(3)_L\times SU(3)_R\times U(1)_{\rm em}\times
U(1)_B\mapsto SU(3)_V\times U(1)_{\tilde Q}\times Z_2,
\end{equation}
where $SU(3)_V$ is the diagonal subgroup of three $SU(3)$ groups and
the generator of $U(1)_{\tilde Q}$ is a linear combination of the
color hypercharge and $U(1)_{\rm em}$ generator,
\begin{equation}
\tilde Q=\cos\theta Q_{\rm em}+\sin\theta Y_8,
\end{equation}
where $\tan\theta=e/g_s$.

\section{Meson mass}
In the CFL phase of color superconductors, there are
8 pseudo Nambu-Goldstone (NG) bosons and one genuine NG boson.
Since the (pseudo) NG bosons are very light, they constitute the
low-lying excitations of the CFL phase, together with the modified
photon, which are relevant in the low energy phenomena
like the cooling process of color superconductors.

The pseudo NG bosons will get mass due to interactions,
that break $SU(3)$ chiral symmetry, such as Dirac mass
terms~\cite{Son:2000cm,Hong:2000ei,Rho:2000xf,Manuel:2000wm,Beane:2000ms}
electromagnetic interactions~\cite{Hong:2000ng,Manuel:2000xt},
and instantons~\cite{Manuel:2000wm}.
It is important to note that Dirac mass term and intanton effects
are suppressed
by powers of $1/\mu$ at high density since they invlove anti-quarks,
while the electromagnetic interaction is not.
In this section we derive the meson mass due to the Dirac mass term
and the electromagnetic interaction, by matching the vacuum energy shift
in microscopic theory (QCD) and the effective theory of mesons, which was
used in~\cite{Son:2000cm,Hong:2000ei}. But, we present the calculation,
using the effective theory constucted in~\cite{Hong:2000tn,Hong:2000ru}, as
was done by Beane {\it et. al}~\cite{Beane:2000ms}.
As in~\cite{Hong:2000tn,Hong:2000ru}, if we introduce the charge conjugated
field $\psi_c=C{\bar\psi}^T$ with $C=i\gamma^0\gamma^2$
and decompose the quark field into
states ($\psi_+$) near the Fermi surface and the states
($\psi_-$) deep in the Dirac sea, the
Dirac mass term can be rewritten as
\begin{equation}
m_q\bar\psi\psi={1\over2}m_q
\left(\bar\psi_+\psi_-+\bar\psi_-\psi_+\right)+
{1\over2}m_q^T
\left(\bar\psi_{c+}\psi_{c-}+\bar\psi_{c-}\psi_{c+}\right),
\end{equation}
which becomes,
if one integrates out the antiquarks ($\psi_-$ or
$\psi_{c-}$),
\begin{eqnarray}
m_q\bar\psi\psi&=&{m_q^2\over2\mu}\psi_+^{\dagger}
\widehat{\psi_-\bar\psi_-}
\psi_+ +
{m_qm_q^T\over2\mu}\psi_+^{\dagger}
\widehat{\psi_{-}\bar\psi_{c-}}
\psi_{c+}+\cdots,
\label{dmass}
\end{eqnarray}
where
$\widehat{\psi_-\bar\psi_-}$ and
$\widehat{\psi_{-}\bar\psi_{c-}}$ are the antiquark propagators,
propagating into the antiquark field itself or its charge-conjugated
field, respectively.
Antiquark fields propagate into their charge conjugated fields only if
they have a Majorana mass and thus the meson mass due to Dirac mass is zero
if the antiquark Majoran mass is zero.

At first one may think that the Majorana mass of antiquarks is zero,
since
it is energetically not prefered for them to develop a condensate
due to the gap ($\sim\mu$) to create an antiquark. But, it is
shown~\cite{Hong:2000ng}
that the antiquark fields get a radiative Majorana mass,
which is equal to the Cooper-pair gap of quarks near the Fermi
surface, since all the symmetries that forbid the Majorana mass term for
antiquarks are broken once the Cooper gap is open for the
quarks near the Fermi surface.

Having shown that the antiquarks have same Majorana mass as quarks near
the Fermi surface,
we may write the inverse propagator of the Nambu-Gorkov
antiquark field, $(\psi_{-},\psi_{c-})^T$, as
\begin{eqnarray}
S^{-1}_c(p)&=&-i{1-\vec \alpha\cdot\vec v_F\over2}
\pmatrix{p_0\gamma_0-\vec p\cdot\vec \gamma+2\mu\gamma_0&
       -{\Delta}^{\dagger} \cr
-{\Delta} & p_0\gamma_0+\vec p\cdot\vec \gamma-2\mu\gamma_0\cr}\\
&=&-i{1-\vec \alpha\cdot\vec v_F\over2}\gamma_0
\pmatrix{l\cdot V+2\mu&
       -{\Delta}^{\dagger} \cr
-{\Delta} & l\cdot\bar V-2\mu\cr},
\end{eqnarray}
where the projector
$(1-\vec\alpha\cdot\vec v_F)/2$ is to project out the
states in the Dirac sea,
$V^{\mu}=(1.\vec v_F)$, ${\bar V}=(1,-\vec v_F)$, and
we decompose $p^{\mu}=\mu v^{\mu}+l^{\mu}$ with
$v^{\mu}=(0,\vec v_F)$ in the second line.
Since the states in the Dirac sea can propagate into their
charge conjugated states via the radiatively generated
Majorana mass term, Eq.~(\ref{dmass}) becomes
\begin{eqnarray}
m_q\bar\psi\psi&=&{m_q^2\over2\mu}\psi_+^{\dagger}\left(
1-{i\partial\cdot V\over2\mu}\right)\psi_+ +
{m_qm_q^T\over4\mu^2}\psi_+^{\dagger}\Delta\psi_{c+}+\cdots,
\end{eqnarray}
where $V^{\mu}=(1,\vec v_F)$ and
the ellipsis denotes the terms higher order in $1/\mu$.
Then, the vacuum energy shift due to the Dirac mass term is
$\sim m_q^2\Delta^2\ln(\mu^2/\Delta^2)$ in the leading order,
which has to be matched with the vacuum energy in the
meson Lagrangian, $m_{\pi}^2F^2$ with the pion decay
constant $F\sim\mu$~\cite{Son:2000cm}.
Therefore, one finds
the meson mass due to the Dirac mass
$m_{\pi}^2\sim m_q^2 \Delta^2/\mu^2\cdot
\ln(\mu^2/\Delta^2)$~\cite{Son:2000cm,Hong:2000ei,Rho:2000xf,Manuel:2000wm,Beane:2000ms}.
The electromagnetic interaction also contributes
to the meson mass, since it breaks the $SU(3)$
flavor symmetry. Among 8 pseudo Nambu-Goldstone
bosons, four of them have the unbrken $U(1)$ charge and
receive a correction,\cite{Hong:2000ng,Manuel:2000xt}~
$\delta m_{\pi}\simeq 12.7\sin\theta\Delta
\left[\ln(\mu^2/\Delta^2)\right]^{1/2}$,
where $\theta=\tan^{-1}\left(e/g\right)$.~\footnote{
In~\cite{Manuel:2000xt}, the result is a little different from the
one obtained in~\cite{Hong:2000ng}
}

Finally, the instanton breaks the chiral symmetry and contributes
to the meson masss. But, its effect at high density
is suppressed by $\mu^{-14}$
for three flavors~\cite{Schafer:2000fe} and thus negligible.

\section{Bosonization of the CFL dense QCD}

In this CFL phase, the
particle spectrum can be precisely mapped into that of the
hadronic phase at low density. Observing this map, Sch\"afer and
Wilczek~\cite{Schafer:1999ef,Schafer:1999pb} further conjectured
that two phases are in fact continuously connected to each other.
The CFL phase at high density is complementary to the hadronic
phase at low density. This conjecture was subsequently
supported~\cite{Hong:1999dk,Hong:2000ff} by showing that quarks in the CFL
phase are realized as  Skyrmions, called superqualitons, just
like baryons are realized as Skyrmions in the hadronic phase.

This phase continuity can be explained heuristically in the following
thought experiment.
Suppose we inject a hydrogen atom into a CFL color superconductor as in
Fig.~3. In the color superconductor,
being bombarded by energetic (gapped) quarks,
the atom will be ionized and the quarks in the proton will
get deconfined to form, for example, a Cooper pair of $u$ and $d$, leaving
the up quark alone.
\begin{figure}[h]
\vskip 0.3in
\epsfxsize=4.0in
\centerline{\epsffile{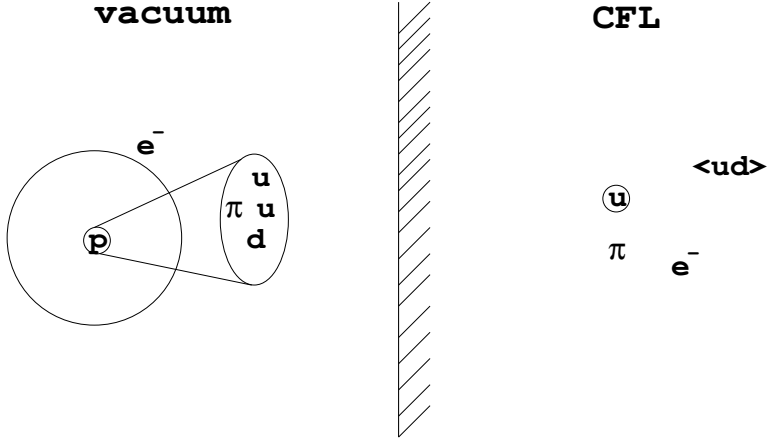}}
\caption{The phase continuity. }
 \label{conti}
\end{figure}
\noindent
From this thought experiment we find two things:
The baryon number of up quark is same as that of
proton, since the CFL vacuum provides the missing two thirds. As the
hydrogen atom is neutral in the vacuum,
the charge of up quark has to be opposite to that of electron with
respect to whatever unbroken charges in the color superconductor.
Thereby, the gapped quarks in the CFL phase correspond to baryons
in the hadronic phase.

Furthermore, it is possible to bosonize the CFL color
superconductor~\cite{Hong:2000ff}, realizing Skyrme's original
motivation for the Skyrmion~\cite{Skyrme:1962vh}.
We introduce a bosonic variable,
\begin{equation}
{U_L}_{ai}(x)\equiv\lim_{y\to x}{\left|x-y\right|^{\gamma_m}
\over\Delta(p_F)}\,\epsilon_{abc}\epsilon_{ijk}
\psi^{bj}_{L}(-\vec v_F,x)\psi^{ck}_{L}(\vec v_F,y),
\end{equation}
where $\gamma_m$ ($\sim\alpha_s$) is the anomalous dimension of the
diquark field.
Similarly, we define $U_R$ in terms of right-handed quarks to describe
the small fluctuations of the condensate of right-handed quarks.
Since the bosonic fields, $U_{L,R}$, are colored, they will interact
with gluons. In fact, the colored massless excitations will
constitute the longitudinal components of gluons through Higgs
mechanism.
Among the small fluctuations of condensates, the colorless
excitations correspond to genuine Nambu-Goldstone (NG) bosons,
which can be described by a color singlet combination of
$U_{L,R}$~\cite{Hong:2000ei,Casalbuoni:1999wu}, given as
\begin{equation}
\Sigma_i^j\equiv U_{Lai}U_R^{*aj}.
\end{equation}
The NG bosons transform under the $SU(3)_L\times SU(3)_R$
chiral symmetry as
\begin{equation}
\Sigma\mapsto g_L\Sigma g_R^{\dagger},\quad {\rm with}\quad
g_{L,R}\in SU(3)_{L,R}.
\end{equation}

Since the chiral symmetry is explicitly broken by the current quark
mass, the instanton effects, and the electromagnetic interaction,
the NG bosons will get mass, which has been calculated by various
groups~\cite{Son:2000cm,Hong:2000ei,Rho:2000xf,Hong:2000ng}.
Here we focus on the meson mass due to the
current strange quark mass ($m_s)$, since it will be dominant for the
intermediate density. Then, the meson mass term is simplified as
\begin{equation}
{\cal L}_m=C\, {\rm tr}(M^T\Sigma)\cdot {\rm tr} (M^*\Sigma^{\dagger})+
O(M^4),
\label{m}
\end{equation}
where $M={\rm diag}(0,0,m_s)$ and
$C\sim \Delta^4/\mu^2\,\cdot\ln(\mu^2/\Delta^2)$. (Note that in general there
will be two more mass terms quadratic in $M$. But, they all
vanish if we neglect the current mass of up and down quarks
and also the small color-sextet component of the Cooper
pair~\cite{Hong:2000ei}.)

Thus, the low-energy effective Lagrangian density
for the bosonic fields in the CFL phase can be written as
\begin{eqnarray}
{\cal L}_{\rm eff}={\cal L}_g
   +   \left[\frac{1}{4}{F}^{2}{\rm tr}(\partial_{\mu}
    U_{L}^{\dag}\partial^{\mu}U_{L})
+n_L{\cal L}_{WZW}
   +(L\leftrightarrow R)\right]+{\cal L}_m
   +g_{s}G_{\mu}^{A}J^{\mu A}+\cdots,
\label{eff_lag}
\end{eqnarray}
where ${\cal L}_g$ is the Lagrangian of Higgsed gluons, $G_{\mu}^A$,
and the ellipsis
denotes the higher order terms in the derivative
expansion, including mixing terms between $U_L$ and $U_R$.
The gluons couple to the bosonic fields through a minimal coupling
with a conserved current, given as
\begin{equation}
J^{A\mu}={i\over 2}F^2{\rm Tr}~U_L^{-1}T^A\partial^{\mu}U_L+
{1\over 24\pi^2}\epsilon^{\mu\nu\rho\sigma}
{\rm Tr}~T^AU_L^{-1}\partial_{\nu}U_LU_L^{-1}\partial_{\rho}U_L
U_L^{-1}\partial_{\sigma}U_L+(L\leftrightarrow R)+\cdots,
\end{equation}
where the ellipsis denotes the currents from the higher order
derivative terms in Eq.~(\ref{eff_lag}).
$F$ is a quantity analogous to the pion decay constant,
calculated to be $F\sim\mu$ in the CFL color
superconductor~\cite{Son:2000cm}. The Wess-Zumino-Witten (WZW)
term~\cite{Wess:1971yu} is described by the action
\begin{equation}
\Gamma_{WZW}\equiv\int{\rm d}^4x\,{\cal L}_{WZW}
=-\frac{i}{240\pi^{2}}\int_{{\sf M}}{\rm d}^{5}r\epsilon^{\mu%
\nu \alpha\beta\gamma}{\rm
tr}(l_{\mu}l_{\nu}l_{\alpha}l_{\beta}l_{\gamma})
\end{equation}
where $l_{\mu}=U_L^{\dag}\partial_{\mu}U_L$
and the integration is defined on a five-dimensional manifold
${\sf M}=V\otimes S^{1}\otimes I$ with the
three dimensional space $V$, the compactified time $S^{1}$, and
a unit interval $I$ needed for the local form of WZW term.
The coefficients of the WZW terms in the effective Lagrangian (\ref{eff_lag})
have been shown to be $n_{L,R}=1$
by matching the flavor anomalies~\cite{Hong:1999dk}, which is
later confirmed by an explicit calculation~\cite{Nowak:2000wa}.

Now, let us try to describe the CFL color superconductor in
terms of the bosonic variables. We start with the effective
Lagrangian (\ref{eff_lag}), which is good at low energy, without
putting in the quark fields. As in the Skyrme model of baryons,
we anticipate the gaped quarks
come out as solitons, made of the bosonic degrees of freedom.
That the Skyrme picture can be realized in the CFL color superconductor
is already shown in~\cite{Hong:1999dk}, but there the mass of the
soliton is not properly calculated. Here, by identifying the
correct ground state of the CFL superconductor in the bosonic
description, we find the superqualitons have same quantum numbers
as quarks with mass of the order of gap, showing that they are
really the gaped quarks in the CFL color superconductor.
Furthermore, upon quantizing the zero modes of the soliton, we find
that high spin excitations of the soliton have energy of order
of $\mu$, way beyond the scale where the effective bosonic
description is applicable, which we interpret as the absence of
high-spin quarks, in agreement with the fermionic description.
It is interesting to note that, as we will see below, by calculating
the soliton mass in the bosonic description, one finds the coupling
and the chemical potential dependence of the Cooper-pair gap,
at least numerically, which gives us a complementary way,
if not better, of estimating the gap.

As the baryon number (or the quark number) is conserved,
though spontaneously broken,
the ground state in the bosonic description should
have the same baryon (or quark) number as the ground state
in the fermionic description.
Under the $U(1)_Q$ quark number symmetry,
the bosonic fields transform as
\begin{equation}
U_{L,R}\mapsto e^{i\theta Q}U_{L,R}e^{-i\theta Q}=e^{2i\theta}U_{L,R},
\end{equation}
where $Q$ is the quark number operator,
given in the bosonic description as
\begin{equation}
Q=i\int {\rm d}^3x~{F^2\over4}
{\rm Tr}\left[U_L^{\dagger}\partial_tU_L-\partial_tU_L^{\dagger}U_L
+\left(L\leftrightarrow R\right)\right],
\end{equation}
neglecting the quark number coming from the WZW term,
since the ground state has no nontrivial topology.
The energy in the bosonic description is
\begin{equation}
E=\int{\rm d}^3x{F^2\over 4}
{\rm Tr}\left[\left|\partial_tU_L\right|^2
+\left|\vec\nabla U_L\right|^2
+\left(L\leftrightarrow R\right)\right]+E_m+\delta E,
\end{equation}
where $E_m$ is the energy due to meson mass and $\delta E$ is the
energy coming from the higher derivative terms. Assuming
the meson mass energy is positive and $E_m +\delta E\ge0$, which
is reasonable because $\Delta/F\ll1$,
we can take, dropping the positive terms due to the spatial derivative,
\begin{equation}
E\ge \int{\rm d}^3x{F^2\over 4}
{\rm Tr}\left[\left|\partial_tU_L\right|^2
+\left(L\leftrightarrow R\right)\right](\equiv E_Q).
\end{equation}
Since for any number $\alpha$
\begin{equation}
\int{\rm d}^3 x~{\rm Tr}\left[\left|U_L+\alpha i\partial_tU_L
\right|^2+\left(L\leftrightarrow R\right)\right]\ge0,
\end{equation}
we get a following Schwartz inequality,
\begin{equation}
Q^2\le  I\,E_Q,
\label{bound}
\end{equation}
where we defined
\begin{equation}
I={F^2\over 4}\int{\rm d}^3x\,{\rm Tr}\left[U_LU_L^{\dagger}
+\left(L\leftrightarrow R\right)\right].
\end{equation}
Note that the lower bound in Eq.~(\ref{bound})
is saturated for $E_Q=\omega Q$ or
\begin{equation}
U_{L,R}=e^{i\omega t} \quad{\rm with}\quad \omega={Q\over I}.
\end{equation}
The ground state of the color superconductor, which has the lowest
energy for a given quark number $Q$, is nothing but a so-called
$Q$-matter, or the interior of a very large
$Q$-ball~\cite{Coleman:1985ki,Hong:1988ur}. Since in the fermionic
description the system has the quark number $Q=\mu^3/\pi^2\int{\rm
d}^3x=\mu^3/\pi^2\cdot I/F^2$, we find, using
$F\simeq0.209\mu$~\cite{Son:2000cm},
\begin{equation}
\omega={1\over\pi^2}\left({\mu\over F}\right)^3 F
\simeq2.32\mu.
\label{fpi}
\end{equation}
In passing, we note that  $\omega$ is numerically very close to
$4\pi F$. 
The ground state of the system in the bosonic description is a
$Q$-matter whose energy per unit quark number is $\omega$.
Now, let us suppose we consider creating a $Q=1$ state out of
the ground state. In the fermionic description,
this corresponds that we excite a gaped quark in the Fermi sea into
a free state, which costs energy at least $2\Delta$.
In the bosonic description, this amounts to creating a superqualiton
out of the $Q$-matter, while reducing the quark number of the
$Q$-matter by one. Therefore, since we gain energy $\omega$
by reducing the quark number of
the $Q$-matter by one, the energy cost to
create a gapped quark from the ground state is
\begin{equation}
\delta {\cal E}=M_Q-\omega,
\label{deltae}
\end{equation}
where $M_Q$ is the energy of the superqualiton configuration.
Since $M_Q\sim 4\pi F$, we see that the energy of the superqualiton
configuration is almost canceled out by $\omega$ to give the gap
much smaller than $4\pi F$ or $\mu$. Varing the strange quark mass, we
find numerically that
the twice of u- and s-superqualiton masses are  given as
\begin{equation}
\begin{array}{lll}
\Delta_{u}=0.079\times 4\pi F,
 &~~~\Delta_{s}=0.081\times 4\pi F, &~~~{\rm for}~m_{K}/F=0.1\\
\Delta_{u}=0.079\times 4\pi F, &
~~~\Delta_{s}=0.089\times 4\pi F, &~~~{\rm for}~m_{K}/F=0.3\\
\Delta_{u}=0.079\times 4\pi F,
&~~~\Delta_{s}=0.109\times 4\pi F, &~~~{\rm for}~m_{K}/F=0.8.
\end{array}
\end{equation}
From the relation that $2\Delta=M_Q-\omega$, we can estimate
numerically the coupling and the chemical potential dependence
of the Cooper gap~\cite{Hong:2000ff}. This gives an alternative
way of calculating the Cooper gap, if not better.

\section{neutrino interaction in CFL}
To discuss the interaction of neutrinos in color
superconductors~\cite{Hong:2000cu},
we first note that gluons mix with weak gauge bosons,
since the diquark condensates in color superconductors carry
not only  a color but also a weak charge.

Consider the color-gluon annihilation into the lepton pair,
$l \bar{l}$, as an example of weak neutral current interactions:
 \begin{eqnarray}
\tilde{V}^{+} + \tilde{V}^- \rightarrow \tilde{Z} \rightarrow \,
\,\,\, l \bar{l}, \label{zgluon}\\
\tilde{V}^{+} + \tilde{V}^- \rightarrow \tilde{V}_0 \rightarrow \,
\,\,\, l \bar{l}. \label{vgluon}
 \end{eqnarray}
The  coupling at the $\tilde{V}\tilde{V}\tilde{Z}$ vertex in the
process mediated by $\tilde{Z}$, Eq.~(\ref{zgluon}), is given by
 \begin{eqnarray}
f \cos^2 \delta\, \frac{4}{\sqrt{3}}\frac{g_s g}{g^2 + g'^2}
\frac{\sigma^2}{v^2} \cos\theta_W \sim \frac{4}{\sqrt{3}} g
\cos^3 \theta_W (\frac{M_V}{M_W})^2 \label{sup}
 \end{eqnarray}
which gives a  suppression factor \begin{eqnarray} \sim  (\frac{M_V}{M_W})^2
 \end{eqnarray}
compared to the conventional $\nu\bar{\nu}$ production.
The suppression  factor in the process mediated by $\tilde{V}_0$,
Eq.~(\ref{vgluon}), due to the vertex $\tilde{V}_0  \,l \bar{l}$ is
given by
 \begin{eqnarray} \sim \sin\beta \,\, \sim \,\, g/g_s. \end{eqnarray}
The propagator in the low energy limit $Q^2 << M_V^2$ is greater
than in Eq.~(\ref{zgluon}), i.e.,
 \begin{eqnarray}
\frac{1}{Q^2 - M_V^2} \sim \frac{1}{M_V^2}. \end{eqnarray}
 However the amplitude for fusion is
enhanced at the strong interaction vertex,
$\tilde{V}\tilde{V}\tilde{V}_0$, by a factor of $f$, and we get
the factor for the amplitude \begin{eqnarray} \sim \, Q_{f} g
\frac{g}{g_s}\frac{1}{M_V^2} g_s = Q_{f}
\frac{g^2}{M_V^2}\label{fusion} \end{eqnarray}
with the modified elctric charge $Q_{f}$.
One can now see that the gluon fusion into the
charged flavor $l \bar{l}$ pair is greater than the weak neutral
current by a factor of $\sim (\frac{M_V}{M_W})^{-2} \sim 10^6$ and
hence comparable to photon mediated processes~\cite{Lee:1978dm}.
However this enhancement does not apply to gluon-mediated $\nu
\bar{\nu}$ processes because $Q_{f}$ is zero for neutrino. In
general, for the neutral current with neutrinos, the contribution
from color-gluon mediated processes in the broken phase vanishes
since the amplitude is proportional to $Q_{f}$(neutrino) which is
$=0$.  We arrive at the same conclusion for $q\bar{q}
\rightarrow  \nu \bar{\nu}$.

The charged current weak interaction
in the process  mediated by $\tilde{V}_0$ is also
comparable to the ordinary weak interaction strength for the
neutrino-quark interaction in the low-energy limit.
Consider the following processes in  matter,
\begin{eqnarray}
q + l &\rightarrow& q' + \nu(\bar{\nu}), \label{qlnu}\\
q &\rightarrow& q' + l + \nu(\bar{\nu}). \label{qnu}
\end{eqnarray}
As in the gluon annihilation processes, there are two amplitudes that
 can be decomposed  into three parts:
 quark gauge boson vertex, propagator, gauge boson-lepton-neutrino vertex,
 \begin{eqnarray}
qq'\tilde{W}^{\pm} \,\,  &\rightarrow& \,  \tilde{W}^{\pm}
\rightarrow  \,\,\,  l\nu\tilde{W}^{\pm}
,  \label{qqz}\\
qq'\tilde{V}^{\pm}\, \,\, &\rightarrow& \,  \tilde{V}^{\pm}\,
\rightarrow \,\,\,  l\nu \tilde{V}^{\pm}. \label{qqv}
 \end{eqnarray}
In the low energy limit, Eq.~(\ref{qqz}) gives the ordinary  weak
amplitude \begin{eqnarray} \sim \frac{g^2}{M_W^2}.\label{qqza}
 \end{eqnarray}
It is easy to see that the contribution of the color
gauge-boson-mediated process, Eq.~(\ref{qqv}), also gives an
amplitude comparable to that of the $W^{\pm}$ mediated process,
\begin{eqnarray} \sim g\frac{g}{g_s}(\frac{M_V}{M_W})^2 \frac{1}{M_V^2}
g_s \sim
\frac{g^2}{M_W^2}.\label{qqva}
 \end{eqnarray}

It should be noted however that the quark decay mediated by
$\tilde{V}_0$ in Eq.~(\ref{qnu}) cannot take place because of the
energy conservation: the quarks with different colors but with
same flavor have the same mass. Therefore the neutrino production
mediated by the color-changing weak current is limited to the
process in Eq.~(\ref{qlnu})
 \begin{eqnarray} q_r \,\, + \,\, e^- \,\,\,
&\rightarrow& \,\,\, q_b \,\, + \nu
\label{nu}\\
q_b \,\, + \,\, e^+ \,\,\, &\rightarrow& \,\,\, q_r \,\, +
\bar{\nu}. \label{nubar}
 \end{eqnarray}
To keep the system in a color-singlet state in the cooling
process, these processes should occur equally to compensate the
color change in each process. It implies that these processes
depend on the abundance of positrons in the system.  At
finite temperature in the cooling period, it is expected that
there will be a substantial amount of positrons as well as electrons
as long as the
temperature is not far below $\sim MeV$. Of course the additional
enhancement of the neutrino production due to the CFL phase
depends on the abundance of positrons in the system which
depends mainly on the temperature. If confined colored gluons are
present in the matter in the CFL phase, the same amplitude can be
obtained in Eq.~(\ref{qqv}) when $qq'$ is replaced with $VV'$.

The result obtained above can be summarized as predicting an
enhancement of the effective four-point coupling constant for the
neutrino production process in the low energy limit.  The
enhancement due to the neutrino-color interaction is suppressed by
factors of $e^{-\Delta/T}$ or $e^{-M_V/T}$, since it depends on
the unpaired excitations above gap which can participate into
neutrino-color interaction. Hence for the cooling process at low
temperatures as $\sim 10^9$K it is not so effective. However
during the early stage of  proto-neutron star the temperature is
expected to be high enough  $\sim 20-50$MeV~\cite{pons} to see the
effect of the enhancement due to color excitations.

Let us now consider the weak decay of
light quasi-quarks, described by the four-Fermi interaction:
\begin{eqnarray}
{\cal L}_{Fermi}&=&{G_F\over\sqrt{2}} \sum_{\vec v_F}
\bar \psi_L(\vec v_F,x)\gamma^{\mu}\psi_L(\vec v_F,x)
\bar \nu_L(x)\gamma_{\mu}\nu_L(x)\\
&=&{G_F\over\sqrt{2}}\sum_{\vec v_F} \psi_L^{\dagger}(\vec
v_F,x){\psi}_{L} (\vec v_F,x) \bar\nu_L(x)\!
\mathrel{\mathop{v\!\!\!\!/}}\nu_L(x)
\end{eqnarray}
where $G_F=1.166\times10^{-5}~{\rm GeV}^{-2}$ is the Fermi
constant and $\psi$ denotes the quasi-quark near the Fermi surface,
projected from the quark field $\Psi$ as in~\cite{Hong:2000ru},
\begin{equation}
\psi(\vec v_F,x)={1+\vec \alpha\cdot \vec v_F\over2} e^{-i\vec
v_F\cdot \vec x}\Psi(x).
\end{equation}

Since the four-Fermi interaction of quarks with opposite momenta
are marginally relevant and gets substantially enhanced at low
energy, it may have significant corrections to the couplings to
quarks of those weakly interacting particles~\cite{Hong:2000ru}:
\begin{eqnarray}
\delta{\cal L}_{\nu q}&=&{G_F\over\sqrt{2}}
\psi^{\dagger}_{L}(\vec v_F,x)\psi_{L}
(\vec v_F,x) \bar\nu_L(x)\!
\mathrel{\mathop{v\!\!\!\!/}}\nu_L(x)\nonumber\\
& &\times {ig_{\bar3}\over 2 M_V^2}\delta^A_{tv;us}\int_y
\left[\bar \psi_t(\vec v_F^{\prime},y)\gamma^0
\psi_s(\vec v_F^{\prime},y)
\bar\psi_v(-\vec v_F^{\prime},y)\gamma^0\psi_u(-\vec v_F^{\prime},y)
\right]\\
&=&{4\over 3}{g_{\bar 3}\over 2\pi}{G_F\over\sqrt{2}}
\psi^{\dagger}_{L}(\vec v_F,x)\psi_{L} (\vec v_F,x)
\bar\nu_L(x)\bar v\cdot\gamma\nu_L(x), \nonumber
\end{eqnarray}
where $\vec v_F$ and $\vec v_F^{\prime}$ are summed over and $g_{\bar3}$
is the value of the marginal four-quark coupling at the screening mass
scale $M$.
In terms of the renormalization group (RG) equation
at a scale $E$
\begin{equation}
{dG_F(t)\over dt}={4\over 3}{g_{\bar 3}(t)\over 2\pi}G_F(t),
\end{equation}
where $t=\ln E$. The scale dependence of the marginal four-quark
coupling in the color anti-triplet channel is calculated
in~\cite{Hong:2000tn,Hong:2000ru}. At $E\ll\mu$
\begin{equation}
{\bar g}_{\bar3}(t)\simeq {4\pi\over 11}\alpha_s(t).
\end{equation}
Since $\alpha_s(t)=2\pi/(11t)$, we get
\begin{equation}
G_F(E)\simeq G_F(\mu)\left({\mu\over E}\right)^{{16\pi\over 363}}.
\end{equation}
Since the RG evolution stops at scales lower than the gap,
the low energy effective Fermi coupling in dense matter is therefore
\begin{equation}
G_F^{\rm eff}=G_F\cdot\left({\mu\over \Delta}\right)^{{16\pi\over 363}}.
\end{equation}
We emphasize that this enhancement applies equally to the $\beta$
decay of quarks and other neutrino production processes described in the
previous section.

At asymptotic density and low temperature ($T\ne0$), the
relevant excitations are quasi-quarks that are not Cooper-paired,
and 17 Nambu-Goldstone bosons. All other massive particles,
Higgsed gluons and other massive excitations are
expected to be out of thermal equilibrium and
decoupled.
Thus the main cooling processes must be the emission of weakly
interacting light particles like neutrinos or other
(weakly interacting) exotic light
particles ({\it e.g.} axions) from the quasi-quarks and
Nambu-Goldstone bosons in the thermal bath.

For the neutrino emissivity
from quasi quarks, the so-called Urca process is relevant.
The neutrino emissivity by
the direct Urca process in quark matter,
which is possible for most cases
in quark matter, was calculated by Iwamoto~\cite{Iwamoto:1980eb}.
For the CFL superconcductor,
we expect the calculation goes in
parallel and the neutrino emissivity is
\begin{equation}
\epsilon_{\rm direct}\propto\alpha_s\rho Y_e^{1/3}T^6,
\end{equation}
where $\rho$ is the density, $T$ is the temperature of
the quark matter, and
$Y_e$ is the ratio between the electron and baryon density.
On the other hand, the neutrino
emissivity by the modified Urca process, which is the dominant
process in the standard cooling of neutron stars~\cite{friman},
is suppressed
by $\left(\Delta/\mu\right)^4$, since the pion coupling to quarks
is given by $g_{qq\pi}\sim \Delta/\mu$~\cite{Hong:2000ei} .
Thus, the neutrino emmisivity by the modified Urca process in the
CFL quark matter is greatly suppressed in the CFL quark matter,
compared to normal quark matter.
Futhermore, since the pion-pion interaction in the CFL quark
matter are also suppressed by
$\Delta/\mu$~\cite{Son:2000cm,Rho:2000xf}, we note that all the
low energy excitations in the CFL quark matter are extremely
weakly coupled. But, since most excitations in the CFL quark matter
are gapped and frozen out, the CFL quark matter has a quite small
heat capacity and cools down very rapidly at temperatures
lower than the gap~\cite{Alford:2000sx}.

 Together with the general
enhancement of the effective four-point coupling constant in RG
analysis,  the enhancement of the neutrino production implies that
the cooling process speeds up as the CFL phase sets in dense
hadronic matter near the critical temperature. But, at temperature
much below the critical temperature, the interaction of quasi-quarks
and pions and kaons is extremely weak, suppressed by
$\Delta/\mu$, and the CFL quark matter cools down extremely rapidly.

For a realistic calculation of the cooling rate of compact stars,
we need to also consider the neutrino propagation in the CFL
matter before the neutrinos come out of the system. A recent
study~\cite{cr} suggests that the presence of the CFL phase can
accelerate the cooling process because neutrino interactions with
matter are reduced in the presence of a superconducting gap
$\Delta$. However this result is subject to modification by the
effect of additional interactions -- not taken into account in
this work -- mediated by the colored gluons on the quark
polarization. It would be interesting to see how the enhancement
of the neutrino production correlates with the neutrino-medium
interaction. This is one of the physically relevant questions on
how the enhanced neutrino interaction could affect
neutron-star(proto neutron star) cooling following supernova
explosion. This issue is currently under investigation.

\section{Conclusion}
I have discussed some aspects of the exciting recent developments in color
superconductivity in high density quark matter.
I have calculated the Cooper pair gap and the critical points
at high density where magnetic gluons are not screened.
The ground state of high density QCD with three light flavors
is shown to be a color-flavor locking state, which can be mapped
into the low-density hadronic phase. The meson mass at the CFL
superconductor is also calculated. The CFL color superconductor
is bosonized, where the Fermi sea is identified as a $Q$-matter and
the gapped quarks as topological excitations, called
superqualitons, of mesons. Finally, as an application of
color supercoductivity, I have discussed the neutrino interactions
in the CFL color superconductor.

I wish to thank the organizers of the 40th Zakopane School of
Theoretical Physics, June 2000 for the
wonderful school, and also the oranizers of
Cosmo-2000, September 2000, Cheju, Korea.
I am grateful to S.-T. Hong, T. Lee,
Y.-J. Park, D.-P. Min, V. Miransky,
M.~Rho, I. Shovkovy,
L.~C.~R.~Wijewardhana,  and I.~Zahed for the collaborations
on the works described here.
This work was supported by  Korea Research Foundation
Grant (KRF-2000-015-DP0069).

%
%
%
%

\end{document}